\documentclass[conference]{IEEEtran}
\usepackage{times}
\usepackage{paralist}
\usepackage{color}
\usepackage{verbatimbox}
\usepackage{blindtext}
\usepackage{tabularx}
\usepackage{url}
\usepackage{graphicx}
\usepackage{fancyvrb}
\usepackage{todonotes}
\usepackage{xcolor}
\DeclareGraphicsExtensions{.pdf,.JPG,.jpg}
\usepackage{listings}
\usepackage{multicol}

\begin{document}
\title{Enhancing the performance of Decoupled Software Pipeline through Backward Slicing}

\author{\IEEEauthorblockN{Esraa Alwan}
\IEEEauthorblockA{Department of Computer Science\\Bath University\\
Email:ehoa20@bath.ac.uk}
\and
\IEEEauthorblockN{John Fitch}
\IEEEauthorblockA{Department of Computer Science\\
Bath University\\
Email: jpff@cs.bath.ac.uk}
\and
\IEEEauthorblockN{Julian Padget}
\IEEEauthorblockA{Department of Computer Science\\
Bath University\\
Email: jap@cs.bath.ac.uk}
}

\maketitle

\begin{abstract}

The rapidly increasing number of cores available in multicore
processors does not necessarily lead directly to a commensurate
increase in performance: programs written in conventional languages,
such as C, need careful restructuring, preferably automatically,
before the benefits can be observed in improved run-times.  Even then,
much depends upon the intrinsic capacity of the original program for
concurrent execution.  The subject of this paper is the performance gains from the combined effect of the complementary techniques of the Decoupled Software Pipeline (DSWP) and (backward) slicing.  DSWP extracts thread-level parallelism from the body of a loop by breaking it into stages which are then executed pipeline style: in effect cutting {\em across} the control chain.  Slicing, on the other hand, cuts the program {\em along} the control chain, teasing out finer threads that depend on different variables (or locations).
parts that depend on different variables.
The main contribution of this paper is to demonstrate that the
application of DSWP, followed by slicing offers notable improvements over DSWP alone,
especially when there is a loop-carried dependence that prevents the
application of the simpler DOALL optimization.
Experimental results show an improvement of a factor of $\approx$1.6 for DSWP + slicing over DSWP alone and a factor of $\approx$2.4 for DSWP + slicing over the original sequential code.


\end{abstract}

\IEEEpeerreviewmaketitle

\vspace{50 cm}
\section{Introduction}
Multicore systems have become a dominant feature in computer architecture.
Chips with 4, 8, and 16 cores are available now and higher core counts are promised.  Unfortunately increasing the
number of cores does not offer a direct path to better performance especially
for single-threaded legacy applications.  But using software
techniques to parallelize the sequential application can raise the
level of gain from multicore systems~\cite{016}.

Parallel programming is not an easy job for the user, who has to deal
with many issues such as dependencies,
synchronization, load balancing, and race conditions.  For this reason
the r\^ole of automatically parallelizing compilers and techniques for the extraction of several threads from single-threaded programs, without
programmer intervention, is becoming more important and may help to deliver
better utilization of modern hardware~\cite{033}.

\begin{figure}[!t]
\begin{center}

\begin{minipage}[b]{0.5\linewidth}

\includegraphics[width=0.9\textwidth]{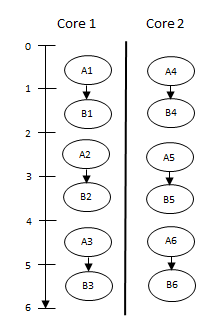}
\end{minipage}
\caption{DOALL Technique adopted from\cite{016}}
\label{fig:d1}
\begin{minipage}[b]{0.43\linewidth}
\includegraphics[width=0.9\textwidth]{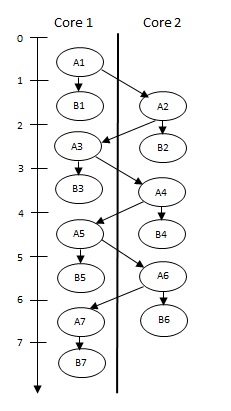}
\label{fig:d2}
\caption{DOACROSS Technique adopted from\cite{016}}
\end{minipage}
\begin{minipage}[b]{0.43\linewidth}
\includegraphics[width=0.9\textwidth]{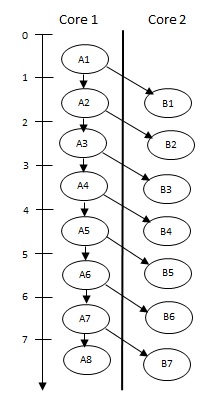}
\label{fig:d3}
\caption{DSWP Technique  adopted from\cite{016}}
\end{minipage}

\end{center}
\end{figure}

Two traditional transformations, whose application typically delivers substantial gains on scientific and numerical codes, are DOALL and DOACROSS. 
DOALL assigns each iteration of the loop to a thread (see figure ~\ref{fig:d1}), which then may all execute in parallel, because there are no cross-dependencies between the iterations.
Clearly, DOALL performance scales linearly with the number of
available threads.  The DOACROSS technique is very similar to DOALL, in that each iteration is assigned to a thread, however, there are cross-iteration data and control dependencies.  Thus, to ensure the correct results, data dependencies have to be respected, typically through synchronization, so that a later iteration receives the correct value from an earlier one as illustrated  in figure (figure ~\ref{fig:d2} ~\cite{016,020}. 
DOALL and DOACROSS  techniques depend on identifying loops that have
a regular pattern\cite{05}, but many applications have irregular
control flow and complex
memory access patterns, making their parallelization very challenging.  The {\bf Decoupled Software Pipeline} (DSWP) has been shown to be an effective technique for the parallelization of applications with such characteristics. This transformation partitions the loop body into a set of stages, ensuring that critical path dependencies are kept local to a stage as shown in figure  ~\ref{fig:d3}. Each stage becomes a thread and data is passed between threads using inter-core communication~\cite{08}. 
The success of DSWP depends on being able to extract the relatively fine-grain parallelism that is present in many applications.

Another technique which offers potential gains in parallelizing general purpose applications is {\bf slicing}.  Program slicing transforms large programs into several smaller ones that execute independently, each consisting of only statements relevant to the computation of certain, so-called, (program) points.  The slicing technique is appropriate for parallel execution on a multi-core processor because it has the ability to decompose the application into independent slices that are executable in parallel~\cite{wang2008new}.

This work explores the possibility of performance benefits arising from a secondary transformation of DSWP stages by slicing.  Our observation is that individual DSWP stages can be parallelized by slicing, leading to an improvement in performance of the longest duration DSWP stages.  In particular, this approach can be applicable in cases where  DOALL is not.

The proposed method is implemented using the {\bf Low level virtual machine}  ({\bf LLVM}) compiler framework~\cite{LLVM}.  LLVM uses a combination of a low level virtual instruction set combined with high level type information. 
An important part of the LLVM design is its intermediate representation (IR).
This has been carefully designed to allow for many traditional analyses and optimizations to be applied to LLVM code and many of which are provided as part of the LLVM framework.
 
The remainder of the paper is organized as follows: the next section (\ref{DSWP}) describes how DSWP may be combined with backward slicing, then section~\ref{Implementation} gives details of the implementation. Section~\ref{results} presents some experimental results from the application of the automatic DSWP + Slicing transformation.  Finally in section~\ref{related work}, we survey related work and conclude (section~\ref{discussion}) with some ideas for future work.


\section{DSWP + Slicing Transformation}\label{DSWP}
The performance of a DSWP-transformed program is limited by the slowest stage.  Thus, any gains must come from improving the performance of that stage.
The main feature of the proposed method is the application of backward slicing to the longest stage emerging from the DSWP transformation.  This is particularly effective when that stage includes a function call.

To illustrate the method, consider the example in Figure~\ref{fig:u}.  DSWP partitions the loop body into the parts labelled L and X, then we slice X to extract  S1 and S2.  Consequently, instead of giving the whole of stage X to one thread, it can be distributed across $n$ threads, depending on the number of slices extracted, with in this case, one core running L (the first stage) and two more running S1 and S2 (the slices from the second stage).

\begin{figure}[t!]
\scriptsize 
\begin{center}
\begin{minipage}[b]{0.4\linewidth}
\begin{lstlisting}

X: Work(cur)
   {
S1:   Slice1(cur);
S2:   Slice2(cur);
   }     
   List *cur = head;
L: for (; cur != NULL; 
          cur = cur->next)   
X:      Work(cur);

\end{lstlisting}
\end{minipage}
\end{center}
\caption{Sliced loop body with recurrence dependency}
\label{fig:u}
\end{figure}

However, while there are potential gains from splitting the loop body into several concurrent threads, there is still the cost of synchronization and communication between threads to take into account.  To minimize these overheads we use lock-free buffers~\cite{09}.  As a result, producer and consumer can access the queue concurrently, via the enqueue and dequeue operations.  This makes it possible for the producer and consumer to operate independently as long as there is at least one data element in the queue.

\section{Implementation of DSWP + Slicing}\label{Implementation}

\begin{figure}[!t]
\hfill\begin{minipage}[t]{0.25\textwidth}
\scriptsize
\begin{lstlisting}[numbers=left]
...
double ss=0;
int i;
double a[0]=0; 
while( node != Null) {(+\label{A}+)
   Calc(node->data,a[i],
    &a[i+1); (+\label{B}+)
   i++;(+\label{C}+)
   node=node->next; (+\label{D}+)
}
...
\end{lstlisting}
\end{minipage}
\hfill%
\begin{minipage}[t]{0.2\textwidth}
\begin{lstlisting}[numbers=left]
Calc(int M,
     double da_in,
     double* da_out) {
   int j;
   b[0]=0;
   for(j=0;j<M;j++) {
      m+=da_in+seq(j);
      (*da_out) +=
         da_in+cos(m);
      b[j]=b[j]+xx(m);
   }
}
\end{lstlisting}   
\end{minipage}
\caption{Source program}\label{fig:source}
\end{figure}
\begin{figure}
\begin{minipage}[t]{0.2\textwidth}
\centering
\includegraphics[width=0.9\textwidth]{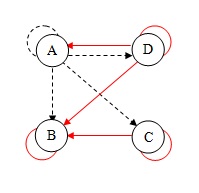}
\caption{Program Dependency Graph }
\label{fig:dd1}
\end{minipage}
\hfill%
\begin{minipage}[t]{0.2\textwidth}
\centering
\includegraphics[width=0.8\textwidth]{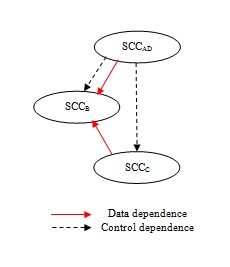}
\caption{DAG of SCCs}\label{fig:dd2}
\end{minipage}
\end{figure}

\begin{figure}[!t]
\begin{minipage}[b]{0.2\textwidth}
\scriptsize
\begin{lstlisting}[numbers=left,xleftmargin=2em]
Slice_1(M,da_in){
 int j;          
 for(j=0;j<M;j++) {
    m+=da_in+seq(j);
    (*da_out) +=
     da_in+cos(m);
  }         
}
\end{lstlisting}
\caption{Slice 1 on {\tt da\_out}}\label{fig:slice 1}
\end{minipage}
\hfill%
\begin{minipage}[b]{0.2\textwidth}
\begin{lstlisting}[numbers=left,xleftmargin=2em]
Slice_2(M,da_in,da_out){
 int j;
 b[0]=0;
 for(j=0;j<M;j++) { 
    m+=da_in+seq(j);
    b[j]=b[j]+xx(m);
  }
}         
\end{lstlisting}
\caption{Slice 2 on {\tt b[j]}}\label{fig:slice 2}
\end{minipage}\hfill
\end{figure}

We build on earlier work by Zhao and Hahnenberg~\cite{035} who implement DSWP in LLVM.  We have extended that code with backward slicing and a decision procedure to determine when it is worth applying the transformation.  The transformation procedure is based on the algorithm for DSWP proposed by Ottoni {\it et al.}~\cite{06}. It takes as input L, the loop to be optimized, and modifies it as a side-effect. The details are as follows:

\begin{compactenum}

\item {\bf Find candidate loop:} This step looks for the most profitable loop to apply DSWP + Slicing.  We collect static information about the program and then use an heuristic to estimate the number of cycles necessary to execute all instructions in every loop in the program.  The loop with the largest estimated cycle count and containing a function call is chosen.  This is a first approximation selection procedure and clearly a more sophisticated version can and should be substituted in due course.

\item {\bf Build the Program Dependency Graph (PDG):} The subject is the loop to be parallelized. Figure ~\ref{fig:dd1} shows that the solid lines (red) denote data dependency and dashed lines (black) control dependency.

\item {\bf  Build strongly connected component (SCC) DAG:} In order to keep all the instructions that   contribute to a dependency local to a thread, a Strongly Connected  Component(SCC) is built, followed by the DAG for the SCCs.  Consider the code in figure~\ref{fig:source}. The loop (lines~\ref{A}--~\ref{D}) traverses a linked list and calls the procedure {\tt Calc}.  
Figure~\ref{fig:dd2} shows the DAGscc of the PDG of the program on the left had side of figure~\ref{fig:source}.  In the procedure {\tt Calc}, there are loop-carried dependencies that make DOALL inapplicable.  DOACROSS is only applicable with the addition of synchronization that may cost more than is gained.
 However, if we can extract independent short slices from this stage and execute them in parallel, the execution time for this long stage can be reduced.  In this case, after DSWP partitioning, we extract two slices (Figures~\ref{fig:slice 1} and~\ref{fig:slice 2}) where function {\tt seq} is side-effect-free.

\item {\bf Assign SCCs to threads:} The previous step may result in more SCCs than available threads.  In this case, we merge SCCs until there are as many as there are threads.
In our example, we have a function call in the loop body. We assign the SCCs that represent the outer loop body to the first thread and the $n$ extracted slices to $n$ threads.

\begin{figure}[h]
\begin{minipage}{0.46\textwidth}
\begin{Verbatim}[fontsize=\small]


Input: A PDG, set of empty list associated, 
one for each node identifier(variable in the
slicing list). 
Output: Slice for each node identifier(variable). 
 
Algorithm:

- Make all PDG nodes as not visited
 
- ComputeASlice(exit node)
 

ComputeASlice ( node n){
     if node is not visited
        Mark node n as visited
        Add the instructions of n to the set
          associated with node n
        For each node m( instruction)in which 
          node n depends  ComputeASlice(m)
        Add the content of the set
          associated with node m to the set
          associated with node n 
                    
}

\end{Verbatim}
\end{minipage}
\caption{The ComputeAllSlice algorithm. Adopted from ~\cite{088}}
\label{fig:algorithm}
\end{figure}
 
\item {\bf Extract slice:} 
In this part, a small slicing program is designed that has the ability to extract slices for the limited range of the case studies.  The algorithm illustrated in figure ~\ref{fig:algorithm} is used to compute an intra-procedural static slice~\cite{088}.
$n$ static slices from the function body are extracted as follows:

In the first step, the PDG is built for the function body by drawing up the dependency table that has both control and data dependency (similar to the one above used to determine thread assignment).
Secondly, the entry block for the function body is examined so as to identify the variables to be sliced and then the names of these are collected, being put on a slicing list. The ComputeASlice is called to extract a slice for every listed variable.
Then, an attempt is made to isolate the control statement parts, such as loop or if statement, into another table called the control table. After collecting the control part instructions, these are added to the extracted slice, if one of the slice instructions is contained in this control parts. For each filtered variable in the slicing identifiers list, first, an  empty list is associated with it and subsequently, all the PDG table entries are scanned to find which one matches the slicing identifier. If one is found, then all the instructions that have data or control dependency are added to the associated list. This procedure is repeated to all the instructions in the associated list and their operands and is not stopped until all the instructions and their operands are contained in this list or all the variables that represent the loop induction variables have been reached. After a set of slices has been extracted from the function body, they are filtered to remove redundant ones so as to avoid repeated calculation, which will happen if all the instructions in one of them have been included in another. For example ,if there are two slices and  slice 1 is completely contained in slice 2 and the second slice (slice 2) is longer than the first, then we will remove the former and keep the latter. This procedure is repeated for all n slices, the real number is obtained.
 In the case of figure~\ref{fig:source} two slice will be retracted for two  variables {\tt da\_out} and {\tt sum}.

\item {\bf Insert synchronization:} To ensure correct results, the dependence between threads must be respected and for pipeline parallelism to be effective, the overhead on core-to-core communication must be as low as possible.  Hence, we use the FastForward circular lock-free queue algorithm~\cite{09}. 
In order to determine the source and the destination of dependencies between the DSWP stages, we need to inspect function arguments. 
These arguments denote the data that will go in the communication buffers.  The destination of a dependency appears in the body of a function and hence where the data must be retrieved in order for the sliced stages to work correctly. 
 
\end{compactenum}   

\section{Experimental Results}\label{results}

This section discusses the results obtained from the application of the automatic implementation of the proposed method that we presented in section~\ref{DSWP}. Several programs have been used as case studies. Some  are artificial and others are taken from~\cite{037}.  The discussion examines two issues:
\begin{inparaenum}[(i)]
\item the effect of lock-free buffers on the performance of DSWP, and
\item the results from the application of DSWP + slicing, demonstrating how this method can improve the performance of long stage DSWP with different program patterns.
\end{inparaenum}

\subsection{Communication Overhead}
This section examines the impact of communication costs on the performance of DSWP. It is important for us to be able to quantify this cost because it is a critical factor in the decision procedure for whether to carry out the DSWP + slicing transformation.  We are also aware this cost will be platform dependent, which is why we provide details of our particular platform.  In a production deployment, this aspect would have to be measured as part of a calibration process.

\begin{figure}[!t]
\scriptsize
\begin{minipage}[b]{0.45\textwidth}
\begin{lstlisting}[numbers=left,xleftmargin=2em]
main()
int N,M
.....
rows=N;
for(i1=1; i1 < rows; i1++) {(+\label{s1}+)
   for(z=1;z<M;z++) {
      sum = 0;
      for(a=1; a<10; a++)
         sum = sum + image[i1]
                    *mask_1[a];       
         if(sum > max) sum = max;
         if(sum < 0)   sum =10;           
         if(sum < out_image[i1])
            out_image[i1] = sum;(+\label{s2}+)
    }
    for(z1=1;z1<M;z1++) {(+\label{s3}+)
       sum1 = 0;
       for(a1=1; a1<10; a1++) {
          sum1 = sum1 + image[i1] 
                      * mask_2[a1];         
          if(sum1 > max) sum1 = max;
          if(sum1 < 0)   sum1 = 10;           
          if(sum1 > out_image[i1])
             out_image[i1] = sum1;(+\label{s4}+)
     }
}
\end{lstlisting}
\caption{Sequential version of program to evaluate DSWP overheads}
\label{fig:fb}
\end{minipage}
\begin{minipage}[b]{0.45\textwidth}
\includegraphics[width=\textwidth]{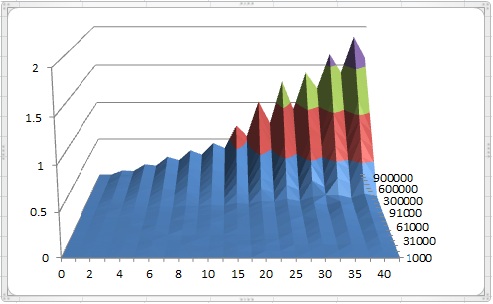}
\caption{Effect of {\tt N and M} on DSWP}
\label{fig:r1}
\end{minipage}
\end{figure}

Consider the program in figure~\ref{fig:fb}. We wish to execute this it by applying  DSWP to the loop that takes the most execution time of the program.

Initially, we partition the program into two parts, give each to a thread and execute the threads as a pipeline. The first thread handles lines~\ref{s1}--\ref{s2} and the second, lines~\ref{s3}--\ref{s4}. Two parameters play a vital role in determining the benefit (or otherwise) of DSWP, namely {\tt M} and {\tt N}.  {\tt M} affects the amount of work inside each thread by controlling the number of iterations in the inner loops, while {\tt N}, in effect, determines the volume of data transfer between threads, by controlling the number of outer loop iterations. 
Figure~\ref{fig:r1} shows how changing the value of {\tt N} (1--40) and {\tt M} (1000--1000000) affects the execution time of the DSWP version compared to the sequential program. From N=6 and M=51000 the performance of DSWP becomes better than the sequential one.

Furthermore the effect of the buffer size on the performance of DSWP is examined, for which the same program as in figure~\ref{fig:fb} was employed. However this time the value of N was fixed to 1,000 and M to 10,000 and the only parameter that was changed was the buffer size. That is, was varied between 10 and 1000, with the execution time of the program being only slightly changed during the during the execution(2 to 5 ms) which was because it was assumed that this was the amount of time needed to create the link list. As a result, it can be concluded that the effect of buffer size on DSWP is trivial.

\subsection{Combining DSWP and slicing}

We now examine the effect of combining DSWP and slicing by applying slicing to the long stage coming out of the DSWP transformation.  The sample programs that we study here all exhibit an imbalance between the two stages of the DSWP, i.e the number of instructions in the outer loop is less than the number of instructions in the function body.  The addition of slicing permits some degree of equilibration. Two of the sample programs are artificial (linkedlist2.c and linkedlist3.c), while the remaining three (fft.c , pro\_2.4.c and test0697.c) are genuine.

For each of the case studies, we extract two slices from the function body, so that the maximum number of threads in general were four depending on whether the extracted slice returns value to the original loop or not.  The data transferred between DSWP stages corresponds to the arguments of a function, which in our case studies are between one and four arguments.

LLVM-gcc (the LLVM C front end, derived from gcc) and the LLVM compiler framework have been used to automate our method. In addition, manually transformed programs have been compiled using gcc in order to be able to compare manual and automatic results.  Table ~\ref{table:0} summarises the technical details of the evaluation platform. 

Our automatic method uses two passes:
\begin{compactenum}
\item The first pass carries out static analysis of all the loops in a program. For each loop it adds up the static execution time for each instruction in the loop body and also accumulates the execution time for the function bodies and stores these results in a table.
\item The second pass chooses a loop to transform and construct the software pipeline. This uses the data collected in the previous pass to identify the highest cost loop, that also contain a function call. 
\end{compactenum}

Next we look at the sample programs in more detail and at the results of the transformation process. 

\begin{table}[!t]
\caption{Platform Details }
\begin{center}
\scriptsize
\begin{tabular}{|l| c| c | c  |}\hline
Processor & Intel(R) Core(TM) i7 CPU  \\ \cline{1-2}           
Processor speed & 2.93 GHz \\\hline
Processor Configuration & 1 CPU, 4 Core, 2 threads per Core \\ \hline
L1d Cache size  & 32 k \\ \hline
L1i Cache size  & 32 k \\ \hline
L2 Cache size  & 256 k \\ \hline
L3 Cache size  & 8192 k \\ \hline
RAM  & 4.GB \\ \hline
Operating System  & SUSE \\ \hline
Compiler & GCC and LLVM \\ \hline
\end{tabular}
\end{center}
\label{table:0}
\end{table}

\begin{figure}[!p]
\begin{minipage}{0.45\textwidth}
  \includegraphics[width=\textwidth]{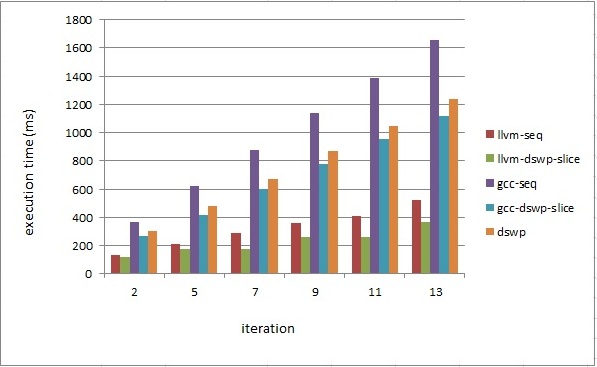}
  \caption{Loop speed up with three threads for test0697.c program}%
  \label{fig:r10}
\end{minipage}\hfill%
\begin{minipage}{0.45\textwidth}
\scriptsize\noindent

\begin{tabularx}{\textwidth}{|X|X|X|X|X|X|}\hline
Iter. & Llvm-seq & Llvm-dswp-slice (Auto.) & Gcc-seq& Gcc-dswp-slice (Man.)&Gcc-dswp \\ \hline\hline
2 &  0.135 & 0.119 & 0.370&0.272&0.304 \\ \cline{1-6}           
5 & 0.215 & 0.173& 0.628 & 0.420&0.483\\ \hline\hline
7 & 0.287 & 0.179 & 0.875& 0.602&0.667\\ \hline\hline
9 & 0.360 & 0.260& 1.140& 0.775&0.866\\ \hline\hline
11 & 0.410 & 0.263 & 1.387&0.954& 1.046 \\ \hline\hline
13 & 0.523 &366 & 1.651& 1.115 & 1.242\\ \hline\hline
\end{tabularx}

  \caption{Execution times for program test0697.c}%
\label{fig:r101}
\end{minipage}
\vfill
\begin{minipage}{0.45\textwidth}
  \includegraphics[width=\textwidth]{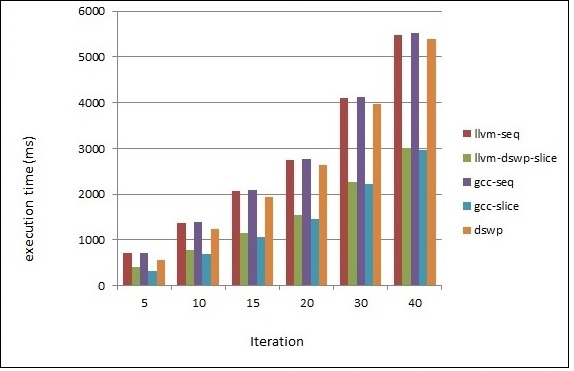}
  \caption{Loop speed up with three threads for fft.c program}%
  \label{fig:r11}
\end{minipage}\hfill%
\begin{minipage}{0.45\textwidth}
\scriptsize\noindent 
\begin{tabularx}{\textwidth}{|X|X|X|X|X|X|}\hline
Iter.&  Llvm-seq& Llvm-dswp-slice (Auto.)&Gcc-seq& Gcc-dswp-slice (Man.) &Gcc-dswp \\ \hline\hline
5  & 0.702 & 0.406&	0.700&	0.310 & 0.558  \\ \cline{1-6}  
10&	1.375&	0.780&	1.391&	0.690 & 1.244\\ \hline\hline
15&	2.058&	1.155&	2.078&	1.069 & 1.934 \\ \hline\hline
20 &2.750&	1.532&	2.770&	1.453 & 2.625\\ \hline\hline
30 &4.106&	2.272&	4.130&	2.214 & 3.972\\ \hline\hline
40 &5.474&	3.013&	5.530&	2.954 & 5.390\\ \hline\hline
\end{tabularx}
  \caption{Execution times for program fft.c}%
  \label{fig:r111}
\end{minipage}
\vfill
  
\end{figure}


{\tt fft.c} An implementation of the fast Fourier transform~\cite{037}.
The test program is a generalization of the program to make it work with N
functions. We give the outer loop to the first thread and the fft function to the second thread.  From the graph in Figure~\ref{fig:r111}, it is clear how the unbalanced long stage DSWP can affect DSWP performance, where it only improves slightly on the sequential program.  We extract two slices from the loop body: the first is the computation of the real part and the second the imaginary part. Figure~\ref{fig:r11} again shows loop speed up for DSWP + Slicing in both manual and automatic forms. 

\begin{figure}[!t]
\begin{minipage}{0.45\textwidth}
  \includegraphics[width=\textwidth]{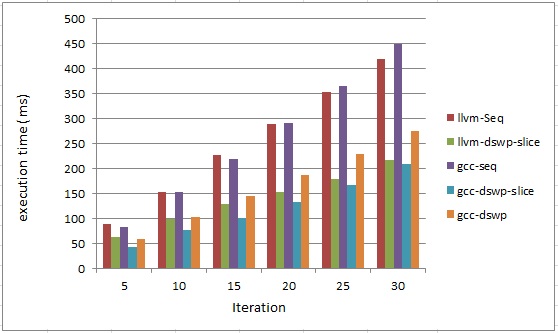}
  \caption{Loop speed up with three threads for Pro\_2.4 program}%
  \label{fig:r12}
\end{minipage}\hfill%
\begin{minipage}{0.45\textwidth}
\scriptsize\noindent 
\begin{tabularx}{\textwidth}{|X|X|X|X|X|X|}\hline
Iter.&  Llvm-seq& Llvm-dswp-slice (Auto.)&Gcc-seq& Gcc-dswp-slice (Man.) &Gcc-dswp \\ \hline\hline
5 &  0.088 &  0.062  &0.83&0.042& 0.058 \\ \cline{1-6}           
10 & 0.153 & 0.100&  0.153 & 0.077&0.103\\ \hline\hline
15 & 0.227 & 0.130  & 0.220& 0.101&0.145 \\ \hline\hline
20 & 0.290 & 0.153& 0.292& 0.134&0.188\\ \hline\hline
25 & 0.353 & 0.180  & 0.365& 0.168 &0.230\\ \hline\hline
30 & 0.419 & 0.217  & 0.450& 0.210 &0.275\\ \hline\hline
\end{tabularx}
  \caption{Execution times for Pro\_2.4 program}%
  \label{fig:r121}
\end{minipage}
\end{figure}

{\tt Pro-2.4.c} This program~\cite{037} 
computes the derivative of N functions. F1 is the first derivative, F2 the second, D1 is the error in F1, and D2 the error in F2. Similar to the previous program we extract two slices from function body after giving the it to the second stage DSWP.  As with the previous program we add  some adaptations to the program and we generalize it to make  it  work for N functions. We set NMAX = 100000 and vary  M from M=5 to M=30. Figure~\ref{fig:r121} shows the execution time for sequential, DSWP, DSWP + slicing (manual) and DSWP + slicing (automatic).   Figure ~\ref{fig:r12} shows loop speed up for Pro\_2.4 using DSWP + Slicing.

{\tt test0697.c}
This program computes the spherical harmonics function, which is used in many physical problems ranging from the computation of atomic electron configuration to the representation of the gravitational and magnetic fields of planetary bodies.
It has two function calls inside the loop body. The first, called the spherical-harmonic-value, gives the initial value to the second function argument, with this function being called the spherical-harmonic. The loop was divided into two parts, depending on the instruction latency execution time. The second function call, which represents the spherical-harmonic was allocated to the second thread, whilst the rest of the loop body containing the first function call was assigned to the first thread. Subsequently, two slices, c[] and s[], were extracted from the second function call by applying slicing technique on this part alone. With high values (40000) of L and M the execution time of this combination was better than for the sequential program. The number of threads was three with two communication buffers and the number of transferred function arguments was four. 
The results obtained by automatic and manual implementation for the sequential and  DSWP\_ Slicing versions, show that the former method gives $\approx$ 1.4  speed up compared with the sequential program in the LLVM environment(see columns 2 and 3 in the table in  ~\ref{fig:r101}).
Moreover, columns 4 and 5 under the GCC environment shows that the speed up becomes $\approx$ 1.5 after applying the slicing technique, while that for DSWP alone is only $\approx$ 1.3.


{\tt linkedlist\{2,3\}.c}  The fourth program is another artificial program in two variants.  The common feature is the traversal of a linked list of linked lists (in contrast to the use of arrays as in the other examples).  The key difference between the variants is that the function called from the loop body does not return a value in the first ({\tt linkedlist2.c}), and does in the second ({\tt linkedlist3.c}).  This allows us to demonstrate the cost of adding a buffer to the program.  Two parameters affect the workload, namely the length of the first level list and the length of the second level list.

\begin{figure}[!t]
\begin{minipage}{0.45\textwidth}
  \includegraphics[width=\textwidth]{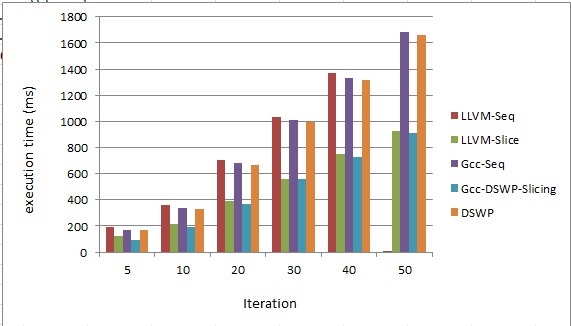}
  \caption{Loop speed up with three threads for linkedlist2.c program}%
  \label{fig:r13}
\end{minipage}\hfill%
\begin{minipage}{0.45\textwidth}
\scriptsize\noindent 

\begin{tabularx}{\textwidth}{|X|X|X|X|X|X|}\hline

Iter.&  Llvm-seq& Llvm-dswp-slice (Auto.)&Gcc-seq& Gcc-dswp-slice (Man.) &Gcc-dswp \\ \hline\hline
5 & 0.191  &0.120  & 0.170 &0.95& 0.167\\ \cline{1-6} 
10 &0.359  & 0.215 & 0.335 & 0.190 & 0.332 \\ \hline\hline         
20 & 0.707  & 0.380 & 0.680 & 0.369 & 0.664  \\ \hline\hline
30 & 1.035  & 0.553  & 1.010 & 0.556 & 0.998 \\ \hline\hline
40 & 1.372 & 0.733 & 1.330 & 0.730 & 1.320  \\ \hline\hline
50 & 1.707& 0.915 & 1.684 & 0.910 & 1.660   \\ \hline\hline
\end{tabularx}
\caption{Execution times for linkedlist2.c program}%
  \label{fig:r131}
\end{minipage}

\begin{minipage}{0.45\textwidth}
  \includegraphics[width=\textwidth]{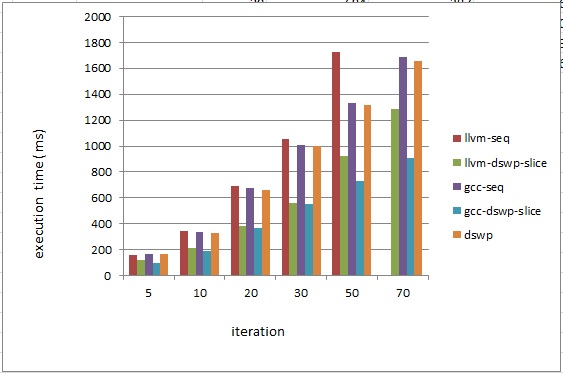}
  \caption{Loop speed up with three threads for linkedlist3.c program}%
  \label{fig:r14}
\end{minipage}\hfill%
\begin{minipage}{0.45\textwidth}
\scriptsize\noindent 

\begin{tabularx}{\textwidth}{|X|X|X|X|X|X|} \hline

Iter.&  Llvm-seq& Llvm-dswp-slice (Auto.)&Gcc-seq& Gcc-dswp-slice (Man.) &Gcc-dswp \\ \hline\hline
5 & 0.160  & 0.122 & 0.170 &0.95& 0.167\\ \cline{1-6} 
10 & 0.344 & 0.214 & 0.335 & 0.190 & 0.332\\ \hline\hline         
20 & 0.694 & 0.387 & 0.680 & 0.369 & 0.664 \\ \hline\hline
30 & 1.058 & 0.557 & 1.010 & 0.556 & 0.998 \\ \hline\hline
50 & 1.726 & 0.927 & 1.330 & 0.730 & 1.320   \\ \hline\hline
70 & 2.440 & 1.286 & 1.684 & 0.910 & 1.660  \\ \hline\hline
\end{tabularx}

  \caption{Execution times for linkedlist3.c program}%
  \label{fig:r141}
\end{minipage}
\end{figure}

In these test the length of the second level list is fixed at 1000 elements, while the length of the first ranges between 10 and 70, giving rise to the results shown in Figure~\ref{fig:r131} and the execution times show in Figure~\ref{fig:r13}.  
  The results for the second version of the program appear in Figure~\ref{fig:r141}.  By comparing Figures~\ref{fig:r131} and~\ref{fig:r141}, we can see how adding an additional buffer to communicate the return value from the one of these slices affects the execution time. This cost appears to have a marginally higher impact on the program using DSWP alone, making it slower than the original sequential program.

\section{Related work}\label{related work}

Weiser\cite{012} proposes the use of slicing for the parallel execution of programs.  He states that slicing is appropriate for parallel execution on multiprocessor architectures,because of the ability to decompose the program into independent slices that execute in parallel without synchronization, or in shared memory by duplicating the computation in each slice.  In general, it is claimed the slices are shorter and execute faster than the original program.  However, there can be an arbitrary difference in the speed of individual slice execution, leading to an interleaving problem ,which is how to find~-- at runtime~-- the correct ordering for slice outputs.  Consequently, after the output of each slice is received, it needs to be reordered to maintain the original program behaviour~\cite{011}.

Wang {\em et al.}~\cite{wang2008new} introduce a dynamic framework to parallelize a single threaded binary program using speculative slicing.  The major contribution of this work can be summarized as: 
\begin{compactitem}
\item Parallelization of binary
 code transparently for multicore systems.
\item Slicing of the `hot' region of the program, rather than the whole program. In addition, they used a loop unrolling transformation that can help to find more loop-level parallelism in a backward slice even in  the presence of loop-carried dependencies and they propose an algorithm to  determine automatically the optimal unrolling factor. They also demonstrate how this factor can affect the parallelism.  
\item Slicing-based parallelism for irreducible control flow graphs.  They define the backward slice using the program dependency graph instead of a program regular expression.  They also introduce the Allow list that uses post-dominator relationships to solve the ambiguity problem that was noted in the previous splicing solution~\cite{011},which is the problem of determining the priority of the instructions in each slice to get the the right output, where the slice output has to be reordered to maintain the original program behaviour.
\end{compactitem}

Rong {\em et al.\/}~\cite{036} propose a method to construct a software pipeline from an arbitrarily deep loop nest, whereas the traditional one is applied to the innermost loop or from the innermost to outer
loops. This approach is called the single-dimensional software pipeline
(SSP).  The (SSP) name came from the conversion of a multi-dimensional data dependency graph (DDG) to 1-D DDG. This approach consists of three steps.
\begin{compactitem}
\item Loop Selection: Every loop level is inspected and  the most profitable one is selected to apply the  software pipeline schedule.  Two criteria can be used to determine which loop is more profitable to the software pipeline schedule are initiation rate and data reuse.
\item  Dependency Simplification: simplify the dependency for the selected loop Lx from the multi-dimension data dependency graph (DDG ) to a single dimension which contains zero dependencies.
\item Final Schedule Computation: after obtaining the simplified DDG, iteration points in the loop nest are allocated to slices: for any i1 in [0,N1], iteration point (i1,0,..,0,0) is assigned to the first slice, (i1,0,..,0,1) to the second, and so on. All i1 iterations can be executed in parallel, if there is no dependency between the iterations and there is unlimited resources. However, if there are dependencies, these iterations will be executed using software pipelines. To address resource limitations, the set of slices are divided into  groups and relegated to succeeding groups until some resources are available. 
\end{compactitem}

Rangan {\em et al.}~\cite{022} introduced a new technique to utilize a
decoupled software pipeline for optimizing the performance of
recursive data structures (RDS) (e.g., linked lists, trees and
graphs). For this kind of structure (RDS),difficulties have been
encountered when trying to execute it in parallel, because the
instructions of a given iteration of a loop depend on the pointer
value that is loaded from a previous iteration. Therefore to address
this problem, a decoupled software pipeline has been used so as to
avoid stalls that are happening with the long variables-latency
instruction in RDS loops. 

 RDS loops consist of two parts, with the first containing the traversal code (critical path of execution) and the second representing the computation that should be carried out on each node traversed by the first part. 
By determining which program part is responsible for the traversal of the recursive data structure, the backward slice for this part should be identified and then decoupled software pipeline techniques can be used to parallelized these parts. The first part will be given to one thread and the second part to another. As the data dependency between these parts is unidirectional (the computation chain in the first part depends on the traversing chain in the second, but not vice-versa) the producer instruction is inserted in the first part and the consumer one in the second.

Raman {\em et al.\/}~\cite{02} introduce a parallel stage decoupled software pipeline (PS-DSWP). This technique is positioned between the decoupled software pipeline and DOALL.  The reason for this combination is that the slowest stage of DSWP bounds the speed of DSWP~-- as we have noted~-- so this work exploits the ability to execute some stages of DSWP using DOALL.  They use special hardware (synchronization array\cite{022}) to communicate data between cores.  For this reason, there is very low communication latency on the performance of PS-DSWP\cite{02}, but the special hardware is experimental and not available on stock processors. 

Huang {\em et al.\/}~\cite{08} show that DSWP can improve performance if it works with other techniques.  This usage called DSWP+, divides the loop body into stages.  These stages are open to parallelization with another techniques like DOALL, LOCALWRITE and SpecDOALL.  After constructing a program dependency graph (PDG) of the loop and finding strongly-connected components (SCCs),the loop body is partitioned into stages.  These stages can be optimized by choosing a suitable parallelizing technique for each stage.  By giving a sufficient number of threads to the parallelization stages, DSWP+ can produce balanced pipelines (there is no big gap in the execution time of the work that is given to each stage). The results suggest that DSWP+ (a combination method) gives more speedup than using DSWP, DOALL, LOCALWRITE alone. It uses lock-free queue and producer and consumer primitives that are implemented in software to communicate data and control condition between threads. LOCALWRITE solves loop carried dependencies for irregular computation over arrays based on array index determination at runtime, however it does not work in all cases.

\section{Conclusion}\label{discussion}

This paper introduces the idea of DSWP applied in conjunction with slicing, by splitting up loops into new loops that are amenable to slicing techniques. An evaluation of this technique on five program codes with a range of dependence patterns leads to considerable performance gains on a core-i7 870 machine with 4-core / 8-threads. The results are obtained from an automatic implementation that shows the proposed method can give a factor of up to 2.4 speed up compared with the original sequential code.

The contribution of this paper is a proof of the concept that DSWP and
slicing can offer useful benefits and, moreover, that such
transformation can be done automatically and under the control of an
heuristic procedure that assesses the potential gains to be achieved.
Consequently, there is much work to be done in respect of improving
the collection of data and the decision procedure, as well as the
integration of the technique into a non-experimental compiler
environment.  More specifically, we aim to increase the potential
parallelism that can be extracted from the long stage DSWP. One of
major issues with backward slice is the longest critical path (slice)
creates a limit on parallelism.  Insights from~\cite{wang2008new}
suggest we can increase parallelism (number of extracted slices) by
combining loop unrolling with backward slice in the presence of loop
carried dependencies.

\section*{Acknowledgment}
 We gratefully acknowledge the Ministry of Higher Education and Scientific Research (MoHESR) in Iraq for their financial support during the period of this research.

\bibliographystyle{plain}
\bibliography{cc2012}

\end{document}